\documentclass[conference]{IEEEtran}
\IEEEoverridecommandlockouts
\usepackage{cite}
\usepackage{amsmath,amssymb,amsfonts}
\usepackage{algorithmic}
\usepackage{graphicx}
\usepackage{textcomp}
\usepackage{xcolor}

\def\BibTeX{{\rm B\kern-.05em{\sc i\kern-.025em b}\kern-.08em
    T\kern-.1667em\lower.7ex\hbox{E}\kern-.125emX}}
\begin{document}

\title{Demystifying  Deep Learning Models for Retinal OCT Disease Classification using Explainable AI\\

}

\author{\IEEEauthorblockN{ Tasnim Sakib Apon}
\IEEEauthorblockA{\textit{Computer Science and Engineering} \\
\textit{BRAC University}\\
Dhaka, Bangladesh \\
sakibapon7@gmail.com}
\and
\IEEEauthorblockN{Mohammad Mahmudul Hasan} 
\IEEEauthorblockA{\textit{Electrical and Electronic Engineering} \\ 
\textit{Islamic University of Technology}\\ 
Dhaka, Bangladesh \\ 
mahmudul18@iut-dhaka.edu} 

\and   
\IEEEauthorblockN{ Abrar Islam}
\IEEEauthorblockA{\textit{Electrical and Electronic Engineering} \\
\textit{Islamic University of Technology}\\
Dhaka, Bangladesh \\
abrarislam@iut-dhaka.edu}

\and
\IEEEauthorblockN{ \centerline {MD. Golam Rabiul Alam}}
\IEEEauthorblockA{\textit{dept. of Computer Science and Engineering} \\
\textit{BRAC University}\\
Dhaka, Bangladesh \\
rabiul.alam@bracu.ac.bd}
}

\maketitle

\begin{abstract}
In the world of medical diagnostics, the adoption of various deep learning techniques is quite common as well as effective, and its statement is equally true when it comes to implementing it into the retina Optical Coherence Tomography (OCT) sector,  but (i)These techniques have the black box characteristics that prevent the medical professionals to completely trust the results generated from them  (ii)Lack of precision of these methods restricts their implementation in clinical and complex cases (iii)The existing works and models on the OCT classification are substantially large and complicated and they require a considerable amount of memory and computational power, reducing the quality of classifiers in real-time applications. To meet these problems, in this paper a self-developed CNN model has been proposed which is comparatively smaller and simpler along with the use of Lime that introduces Explainable AI to the study and helps to increase the interpretability of the model. This addition will be an asset to the medical experts for getting major and detailed information and will help them in making final decisions and will also reduce the opacity and vulnerability of the conventional deep learning models. 
\end{abstract}


\section{Introduction}\label{intro}
Explainability is one of the most extensively discussed topics when it comes to the application of artificial intelligence (AI) in healthcare. Despite the fact that AI-driven algorithms have been demonstrated to surpass humans in certain analytical tasks, the absence of explanation has sparked debate. Explainability is more than a technological issue; it also poses a slew of medical, legal, ethical, and social issues that must be thoroughly investigated. In many situations, understanding why a machine learning model produces a certain prediction is just as important as the accuracy of the forecast. Previous studies related to optical coherence tomography (OCT) disease classification have shown excellent results however due to the lack of AI's explainability medical practitioners do not tends to rely on the developed system Since the consequences of any system's judgment might be severe if it is not anticipated accurately. \par

Eye diseases or ocular dysfunctions are fairly frequent in the general population, particularly among the elderly. The most significant anomalies in the eyes are choroidal neovascularization (CNV), diabetic macular edema (DME), and drusen accumulation in the macular region. In the recent decade, optical coherence tomography (OCT) has become one of the most rapidly evolving medical imaging technologies which can capture blood flow, polarization state, structural data, elastic properties, and molecular content, among other things, in biological tissues \cite{optical_co}. Optical diffraction and absorbance of biological tissues can be accurately measured with OCT and thus some diagnostic measures benefit from it. \par
Usually, OCT disease classification is based on the medical practitioners' expertise, which can be inaccurate sometimes. Misdiagnosis and inadequate diagnostic effectiveness can also result from a vision screening. Workload can sometimes result in serious misdiagnosis and the overall process is slower. Our study's overarching objectives are as follows (i) Automating the oct classification procedure. (ii) Reduce the workload of medical professionals. (iii) Increase the efficiency of the entire process. (iv) Provide medical practitioners with specific information so that they may trust our developed system.  \par
This study's contribution may be summarized as follows:
\begin{itemize}
    \item A CNN model is proposed to identify optical coherence tomography diseases in four classes.
    \item In terms of model size, the suggested approach is significantly smaller, making it appropriate for usage as a web app to generate real-time OCT diagnostic classification.
    \item The model is efficient in terms of memory resources,  as well as computational power where model size is less than 2 MB and parameters for training is 423,460.
    \item Using LIME we examine the interpretability of the proposed model, which can provide crucial insights to medical professionals for correct categorization of optical coherence tomography diseases.
    \item Using Grad-CAM we further explore the interpretability to offer important information to medical practitioners.
\end{itemize}
A brief overview of previous OCT disease classification research is included in Section \ref{relatedwork} of this paper, followed by a brief discussion of our methodology, models, and techniques in Section \ref{method}, which is divided into three sub-sections. The explanation of the system model is provided in section \ref{system}, whereas \ref{data} discusses data gathering and preparation procedures, and \ref{model} exposes our proposed CNN model. The performance measurements and results, as well as confusion matrices, have been depicted in Section \ref{performance}. Finally, in section \ref{explain} we interpret our model and attempt to illustrate how it makes decisions.
\begin{figure}[!tbp]
  \centering
  \begin{minipage}[b]{0.23\textwidth}
    \includegraphics[width=\textwidth]{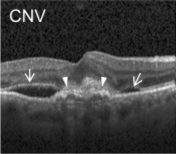}
    \centering {A}
    \label{fig:x CNV}
  \end{minipage}
  \hfill
  \begin{minipage}[b]{0.24\textwidth}
    \includegraphics[width=\textwidth]{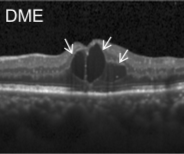}
    \centering {B}
    \label{fig:x DME}
  \end{minipage}
  \hfill

  \begin{minipage}[b]{0.24\textwidth}
    \includegraphics[width=\textwidth]{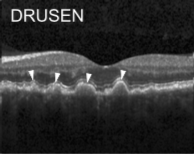}
    \centering {C}
    \label{fig:x DRUSEN}
  \end{minipage}
  \hfill
  \begin{minipage}[b]{0.24\textwidth}
    \includegraphics[width=\textwidth]{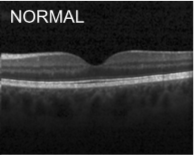}
    \centering {D}
    \label{fig:x NORMAL}
  \end{minipage}
    \caption{Representative Optical Coherence Tomography (OCT) Images (a) CNV  (b) DME (c) DRUSEN (d) NORMAL }
\end{figure}
\section{Related Work}\label{relatedwork}
Various studies have been conducted from different approaches on retina OCT starting from several kinds and shapes of deep neural network and currently applying the explainable AI method. Timo ET Al. has developed a CNN model that can segment the total retinal scans area that has been produced using a self-full-filed OCT which is comparatively cheaper as well as can also segment the Pigment Epithelial Detachments (PED) and observed that retina segmentation provides a considerably better result than that of PED (0.939 and 0.6 DSC score respectively) \cite{kepp2020segmentation}. Moreover, Jing ET Al. presented a quality assessment system that can automatically classify OCT images based on factors like signal completeness, location and effectiveness i.e., assessing that image’s quality and using this technique, he filtered out data and achieved an increase of accuracy by 3.75\% \cite{wang2019deep}. 
Recently, experiments and researches on Retina OCT have been conducted based on Explainable AI, as the conventional deep learning approach may have an adequate value in terms of accuracy, though it lacks information when it comes to the explanation of this result, which is significant especially in the medical field.  In this respect, Aniket ET. Al. presented an approach by providing a CAD model that uses class labels and rough localization for acquiring information and can find out the evidence before diagnosis and detects diabetic macular edema (DME) from OCT slices as an illustration \cite{joshi2020explainable}. In addition, Amitojdeep ET. Al. has illustrated a comparative analysis for the purpose of identifying the best explanations methods in terms of retinal OCT diagnosis \cite{singh2020quantitative}. 13 different attribution methods have been used and the Deep Taylor attribution method got the highest rating providing a median rating of 3.85/5. 

\begin{table}[!ht]
\begin{center}
\caption{Related Works.}
    \scalebox{1}
    {
        \begin{tabular}{|c |c |c |c |c| c} 
        \hline
        Architectures  & Test Accuracy &  Sensitivity &  Specificity & Ref\\ [0.9ex] 
        \hline
        Lee et al.  & 87.63 & 84.63  & 91.54 & \cite{lee}\\
        \hline
        Human Expert 2  & 92.10 & 99.39 & 94.03  & \cite{human_expart} \\ 
        \hline
        Awais et al.  & 93.00 & 87.00 & 100.00 & \cite{awais}\\
        \hline
        ResNet50-v1 & 94.92  & 94.92 & 97.46 & \cite{relatedworkresnet}\\
        \hline

        \end{tabular}
        \label{relatedworks}
    }
\end{center}

\end{table}
Unlike the previous studies, we concentrated on making our model smaller and faster which will be efficient in real-time or for web app, and later focused on explaining the model.
\section{Methodology}\label{method}
We obtain a clear overview of our proposed model, which is separated into three subsections, from Section \ref{method}. Part \ref{system} discusses our system model, followed by part \ref{data}, which discusses data gathering and pre-processing, and lastly, part \ref{model}, which discusses our pre-trained and self-made deep neural network models.

\subsection{Proposed Retinal OCT Model}\label{system}
Our dataset, containing 84,495 X-ray images of four different categories including Normal, CNV, DME, DRUSEN \cite{Kermany2018}. For this huge dataset to proceed with the CNN model there is a resource-constrained. So, data is handled differently without preloading all images. Depending on the subfolders, each image is labeled accordingly.  Labels and names of images are stored in two different variables. They are partitioned into two different disjoint sets as train and test after shuffling. A custom image generator works in the middle to produce a batch of image files using labels and names. The batch size can easily be manipulated based on available resources. These batches of images are sequentially fed into a custom-made 6-layer CNN model. Testing data is employed to evaluate the performance. This model is then explained through LIME (Local Interpretable Model-agnostic Explanations), an explainable AI. This explanation can help to build trust among doctors to use CNN model as a solution in the medical field. However, it can also unveil hidden insight behind the model prediction.
\begin{figure}[!ht]
\centering
\includegraphics[scale=0.65]{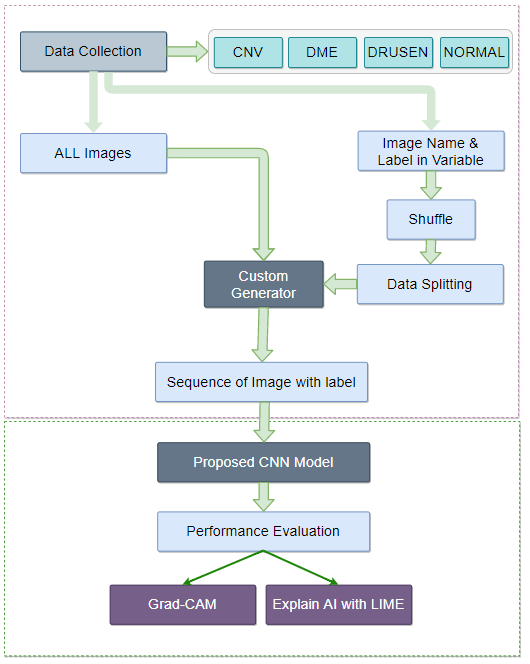}
\caption{Proposed Memory Efficient System.}
\label{fig:x system}
\end{figure}
\subsection{Data Acquisition and Preparation } \label{data}
Optical coherence tomography (OCT) images were selected from retrospective cohorts of adult patients from different hospitals around the world. OCT Images are labeled as (disease)-(randomized patient ID)-(image number by this patient) \cite{Kermany2018} and Fig 1 shows some examples of OCT images. Image properties are presented in Table II and the distribution of image quantity is presented in Table III. Custom image generators resize images to 224x224x3 and normalize them


\begin{table}[!ht]
\begin{center}
\caption{Data Details.}
    \scalebox{1}
    {
        \begin{tabular}{|c |c |c |c |c| c} 
        \hline
        Image Class  & Min Width &  Min Height &  Max Width & Max Height\\ [0.9ex] 
        \hline
        CNV  & 512 & 496 & 512 & 496\\ 
        \hline
        DME  & 512 & 496 & 512 & 496\\
        \hline
        DRUSEN  & 512 & 496 & 768 & 496\\
        \hline
        NORMAL  & 512 & 496 & 768 & 512\\
        \hline
        \end{tabular}

        \label{datadetails}
    }
\end{center}

\end{table}

\begin{table}[!ht]
\begin{center}
\caption{Distribution of Data.}
    \scalebox{1.1}
    {
        \begin{tabular}{|c |c |c |c |c| c} 
        \hline
        Image Class & Available & Training & Validation &  Testing\\ [0.5ex] 
        \hline
        CNV & 37742 & 22646 & 7548 & 7548\\ 
        \hline
        DME & 11840 & 7104 & 2368 & 2368 \\
        \hline
        DRUSEN & 9108 & 5466 & 1821 & 1821 \\
        \hline
        NORMAL & 26807 & 16085 & 5361 & 5361 \\
        \hline
        \end{tabular}
        \label{fig:x distribution}
    }
\end{center}

\end{table}

\subsection{Model Specification} \label{model}
This model consisting of only 6-layer model contains 423,460 parameters for training. Four consecutive convolutional layers following two fully connected layers with ReLU activation function are embraced in this model. Filter size is used sequentially 64-32-128-128 for convolution layers possessing strides 2-1-1-1 on each dimension. Output is filtered by connected layers activated with sigmoid function. Each 3x3 max-pooling layer bundled with a convolution layer helps significantly reducing the function. The model network is shown in Fig 3 and the model configuration in Table IV.

\begin{figure}[!ht]
\centering
\includegraphics[scale=0.73]{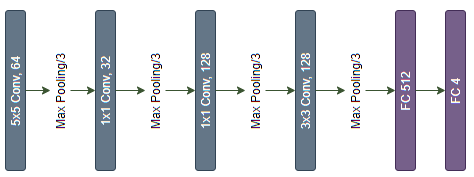}
\caption{Proposed 6-layer System Architecture.}
\label{fig:x data}
\end{figure}
\par Trained model is then used for prediction and LIME is introduced as the explain-er for this model. 100 samples are used with 5 top-levels to explain the model as there is 
\begin{table}[!ht]
\begin{center}
\caption{System Model.}
  
        \begin{tabular}{|c |c |c |c |c} 
        \hline
        Layer  & Input Shape &  Output Shape &  Param \\ [0.9ex] 
        \hline
        Conv5D-64 & 224x224x3 & 110x110x64 & 4864 \\ 
        \hline
        Max Pooling  & 110x110x64 & 54x54x64 & 0 \\
        \hline
        Conv1D-32 & 54x54x64 & 54x54x32 & 2080 \\
        \hline
        Max Pooling  & 54x54x32 & 26x26x32 & 0 \\
        \hline
        Conv1D-128 & 26x26x32 & 26x26x128 & 4224 \\
        \hline
        Max Pooling  & 26x26x128 & 12x12x128 & 0 \\
        \hline
        Conv3D-128 &  12x12x128  &  12x12x128 & 147584 \\
        \hline
        Max Pooling  & 12x12x128 & 5x5x128 & 0 \\
        \hline
        Dropout & 5x5x128 & 5x5x128 & 0 \\
        \hline
        Max Pooling  & 5x5x128 & 2x2x128 & 0 \\
        \hline
        Flatten & 2x2x128 & 512 & 0 \\
        \hline
        Dense 1 & 512 & 512 & 262656 \\
        \hline
        Dropout & 512 & 512 & 0 \\
        \hline
        Dense 2 & 512 & 4 & 2052 \\
        \hline
        & {Total Param}  & 423460 &\\
        \hline
        & {Total Trainable Param}  & 423460 &\\
        \hline
        \end{tabular}

        \label{fig:x model}
    
\end{center}

\end{table}

\section{Performance Evaluation} \label{performance}

\begin{figure}[!tbp]
  \centering
  \begin{minipage}[b]{0.48\textwidth}
    \includegraphics[width=\textwidth]{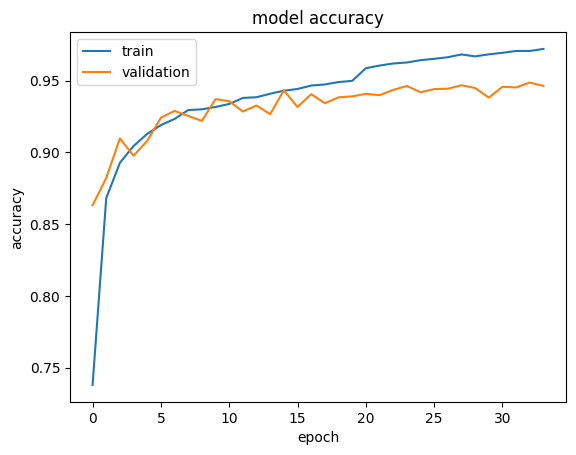}
    \centering {A}
    \label{fig:x ac}
  \end{minipage}
  \hfill
  \begin{minipage}[b]{0.48\textwidth}
    \includegraphics[width=\textwidth]{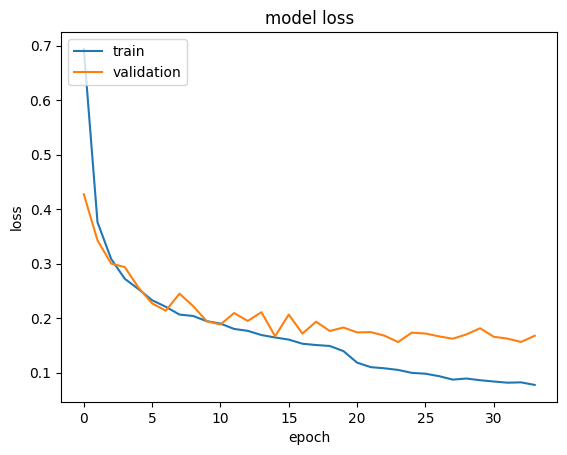}
    \centering {B}
    \label{fig:x loss}
  \end{minipage}
  \hfill
    \caption{Model Accuracy \& Loss.}
\end{figure}

From Section \ref{performance} we get an overview of our proposed system's performance. \par
The dataset has been trained using our custom-made model dividing the whole data set into a batch size of 64 and later a portion of the dataset that has been extracted before the training as the test dataset and has been used in order to measure the performance of the model using various performance metrics. In addition to our custom model, the performance of the pre-trained model VGG-19 on the same dataset has been determined to get a clear comparative idea between the custom model and the pre-trained model. Accuracy of the overall dataset and precision, F-1 Score and sensitivity of each of the classes for each of the models has been calculated as the measurement of performance evaluation and has been depicted in Table V. \par

Firstly, if we focus on the models, the self-developed custom model has a clear superiority in terms of performance in all the performance metric. VGG-19 achieves an accuracy of 92.77\% on the overall dataset, whereas the custom model reaches 94.87\% exceeding that of VGG-19. Not only accuracy but also in terms of other metrics, the custom model always gains a higher value than the VGG-19 model in any of the classes. The precision, F-1 Score, sensitivity values of 0.921, 0.928 and 0.921 respectively of the CNV class from the custom model compared to the 0.898, 0.917 and 0.898 values from VGG-19 model can be given as an example that infers the previous statement. Fig 5 illustrates the confusion matrix of all the labels which is produced from the result of the proposed model \par

Secondly, another important observation from Table V is among the four classes CNV and Normal class has been comparatively superior in case of all the metrics, as the precision, F1-score and recall of CNV from the custom model are 0.967,  0.966 and 0.967 respectively and that values for the Normal cases are 0.977, 0.964  and 0.977 and similar scenario for the pre-trained model. The reason behind their better performance is the presence of their larger number of data and as the larger, the number of the data, the better the performance of any CNN model would be. The latter statement is also proven by the fact that the Drusen class has provided the poorest result and the quantity of data for this class is also the smallest So, in the end, it can be inferred that the performance of the self-developed custom model, as well as the pre-trained model, can easily be upgraded by collecting more data.
\par Overall, in terms of computation 6-layer custom model is more efficient than VGG-19. The parameters to be trained in our model is 423,460 wherein VGG-19  it is 21,011,740. Again, the model weight size of 6-layer custom model is less than 2 MB having a clear edge over the VGG-19.

\begin{figure}[!t]
\centering
\includegraphics[scale=0.325]{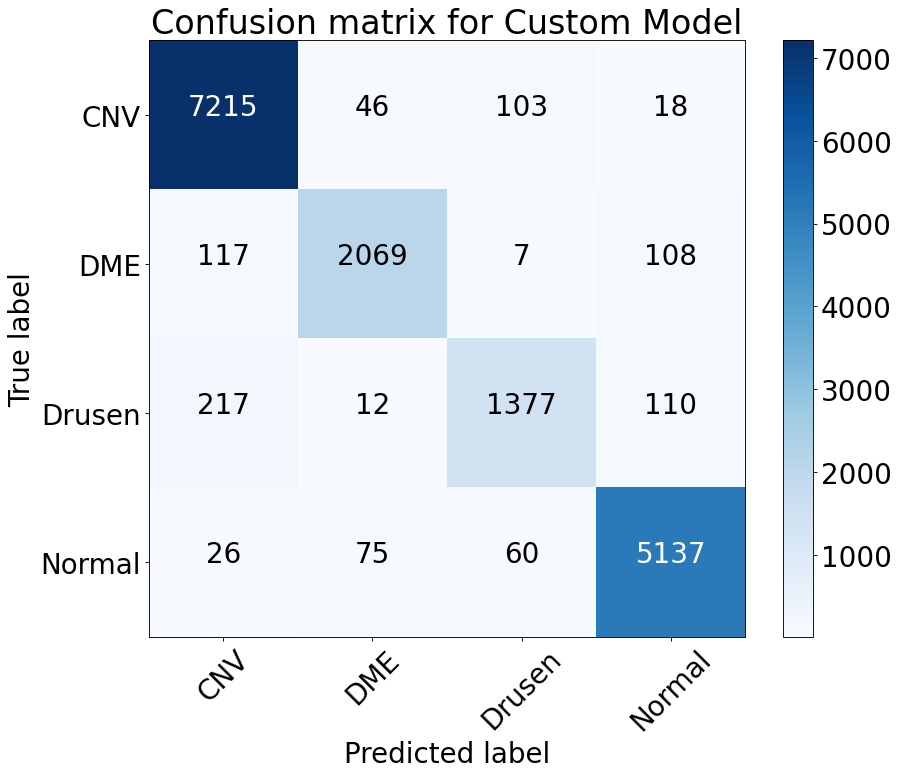}
\caption{Confusion Matrix of the Proposed Model.}
\label{fig:x confusion}
\end{figure}

\begin{table}[!http]
\begin{center}
\caption{Model Performance.}

\begin{tabular}{|c|c|c|c|c|c| } 
\hline
Model & Action & Precision & F-1 Score & Sensitivity & Accuracy\\
\hline \hline
 & CNV &  0.967 & 0.966 & 0.967 & \\ 
Custom & DME & 0.921  & 0.928 & 0.921 & {94.87}\\ 
& Drusen & 0.825  &  0.852 & 0.825 &\\
& Normal & 0.977  & 0.964 & 0.977 &\\ 
\hline
& CNV & 0.964  & 0.962 & 0.964 & \\ 
VGG-19 & DME & 0.898 & 0.917 & 0.898 & {92.77}\\ 
& Drusen & 0.832 & 0.829 & 0.832 & \\ 
& Normal & 0.953 & 0.948 & 0.953 &\\ 
\hline

\hline
\end{tabular}
\label{modelscore}
\end{center}

\end{table}

\begin{figure*}[!t]
	\centering
	\begin{minipage}{.4\columnwidth}
		\centering
		\includegraphics[width=\textwidth]{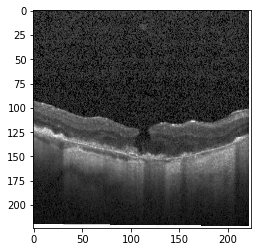}
		{Original Image}
		\label{label11}
	\end{minipage}%
	\begin{minipage}{.4\columnwidth}
		\centering
		\includegraphics[width=\textwidth]{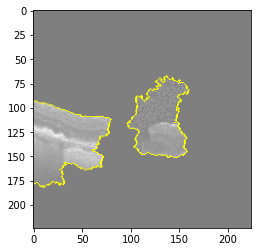}
		{Explanation B}
		\label{image_b4}
	\end{minipage}
	\begin{minipage}{.4\columnwidth}
		\centering
		\includegraphics[width=\textwidth]{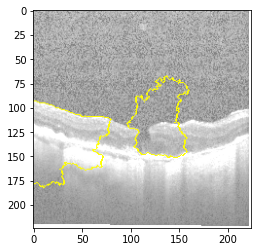}
		{Explanation C}
		\label{image_c}
	\end{minipage}
	\begin{minipage}{.4\columnwidth}
		\centering
		\includegraphics[width=\textwidth]{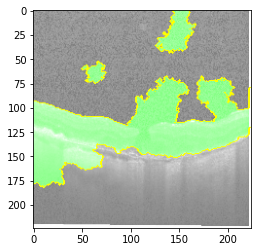}
		{Explanation D}
		\label{label21}
	\end{minipage}
	\begin{minipage}{.37\columnwidth}
		\centering
		\includegraphics[width=\textwidth]{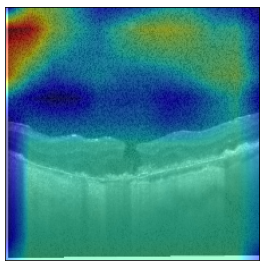}
		{GradCAM}
		\label{label22}
	\end{minipage}

	\caption{Visualization of correct prediction, Class : CNV.}
	\label{XAI_CNV}
\end{figure*}

\begin{figure*}[!t]
	\centering
	\begin{minipage}{.4\columnwidth}
		\centering
		\includegraphics[width=\textwidth]{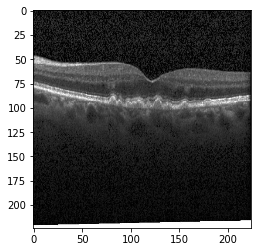}
		{Original Image}
		\label{label10}
	\end{minipage}%
	\begin{minipage}{.4\columnwidth}
		\centering
		\includegraphics[width=\textwidth]{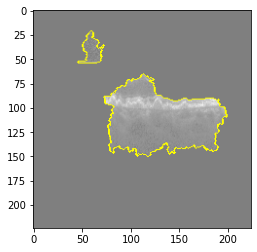}
		{Explanation B}
		\label{image_b1}
	\end{minipage}
	\begin{minipage}{.4\columnwidth}
		\centering
		\includegraphics[width=\textwidth]{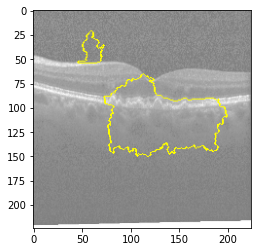}
		{Explanation C}
		\label{label23}
	\end{minipage}
	\begin{minipage}{.4\columnwidth}
		\centering
		\includegraphics[width=\textwidth]{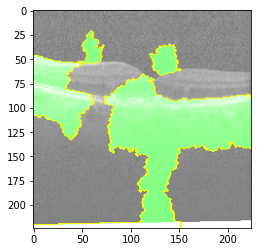}
		{Explanation D}
		\label{label24}
	\end{minipage}
	\begin{minipage}{.37\columnwidth}
		\centering
		\includegraphics[width=\textwidth]{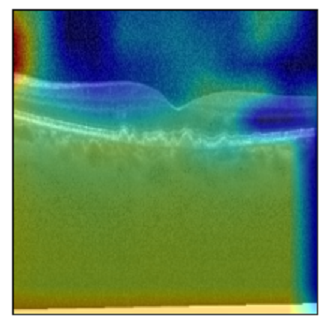}
		{GradCAM}
		\label{label25}
	\end{minipage}

	\caption{Visualization of correct prediction, Class : DRUSEN.}
	\label{XAI_DRUSEN}
\end{figure*}

\begin{figure*}[!t]
	\centering
	\begin{minipage}{.4\columnwidth}
		\centering
		\includegraphics[width=\textwidth]{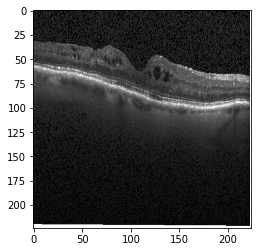}
		{Original Image}
		\label{label12}
	\end{minipage}%
	\begin{minipage}{.4\columnwidth}
		\centering
		\includegraphics[width=\textwidth]{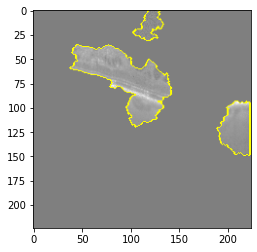}
		{Explanation B}
		\label{image_b2}
	\end{minipage}
	\begin{minipage}{.4\columnwidth}
		\centering
		\includegraphics[width=\textwidth]{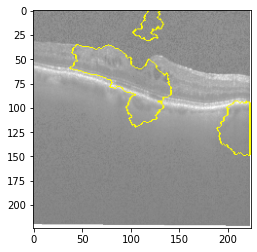}
		{Explanation C}
		\label{label26}
	\end{minipage}
	\begin{minipage}{.4\columnwidth}
		\centering
		\includegraphics[width=\textwidth]{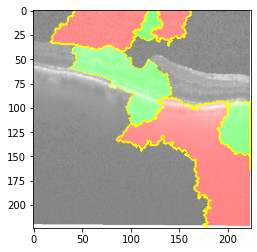}
		{Explanation D}
		\label{label27}
	\end{minipage}
	\begin{minipage}{.37\columnwidth}
		\centering
		\includegraphics[width=\textwidth]{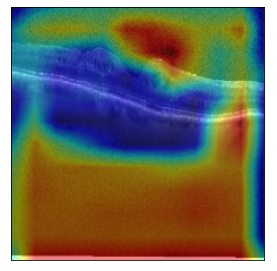}
		{GradCAM}
		\label{label28}
	\end{minipage}

	\caption{Visualization of correct prediction, Class : DME.}
	\label{XAI_DME}
\end{figure*}

\begin{figure*}[!t]
	\centering
	\begin{minipage}{.4\columnwidth}
		\centering
		\includegraphics[width=\textwidth]{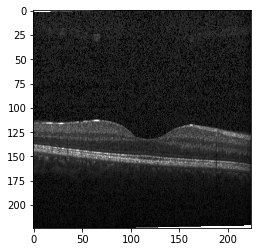}
		{Original Image}
		\label{image_b3}
	\end{minipage}%
	\begin{minipage}{.4\columnwidth}
		\centering
		\includegraphics[width=\textwidth]{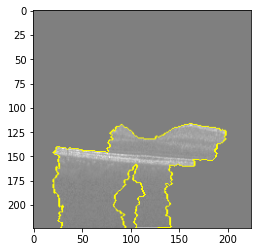}
		{Explanation B}
		\label{label29}
	\end{minipage}
	\begin{minipage}{.4\columnwidth}
		\centering
		\includegraphics[width=\textwidth]{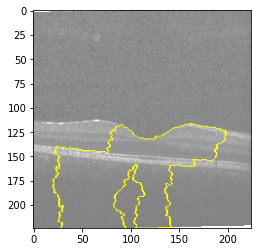}
		{Explanation C}
		\label{label30}
	\end{minipage}
	\begin{minipage}{.4\columnwidth}
		\centering
		\includegraphics[width=\textwidth]{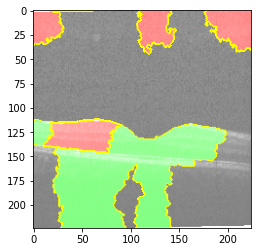}
		{Explanation D}
		\label{label31}
	\end{minipage}
	\begin{minipage}{.37\columnwidth}
		\centering
		\includegraphics[width=\textwidth]{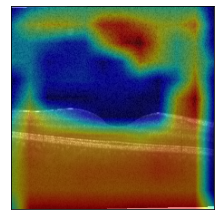}
		{GradCAM}
		\label{label32}
	\end{minipage}

	\caption{Visualization of correct prediction, Class : NORMAL.}
	\label{XAI_Normal}
\end{figure*}

\section{MODEL EXPLANATION WITH XAI} \label{explain}
In this part, we explain the prediction of our proposed system. \par

Fig \ref{XAI_CNV}, \ref{XAI_DRUSEN}, \ref{XAI_DME} and \ref{XAI_Normal} depicts the visualization of correct predictions by our proposed CNN model where fig \ref{XAI_CNV} is class CNV, fig \ref{XAI_DME} is class DME, fig \ref{XAI_DRUSEN} is DRUSEN and finally, fig \ref{XAI_Normal} is NORMAL. Here the first photo in every class is the original image. The LIME map of our suggested model's prediction is shown in image B whereas in image C the positive region is highlighted in specific sections on the original image. For Image D we have increased the number of features from 5 to 10 thus more regions have been predicted as the positive region which is highlighted in green. After increasing the features from 5 to 10, some of the regions are predicted wrongly. The red regions represent the output of incorrect prediction. The following image represents the Grad-CAM heatmap highlighting the regions with our model's prediction.

\section{Conclusion}
A substantial amount of progress has been done in exposing the black-box characteristics of deep learning models. Among the various techniques of Explainable AI, Lime is one of the most widely used for its effectiveness and thus, in this paper, we have examined the explainable capability of the lime on the decisions of our self-developed CNN model in classifying retina OCT diseases. The accuracy achieved by our self-developed model is 94.87\% which is satisfactory, however in case of making critical and crucial decisions, even a single error can cost badly and hence to mitigate that problem the application of Lime AI comes, which will assist the medical professionals to analyze the predicted results from the model and over-turn the result where it is necessary. Furthermore, our proposed approach is simpler and requires comparatively less amount of computational power and memory, therefore it should be well-suited in real-time application. Hopefully, the application of Explainable AI will help to reduce the uncertainty of deep learning models and thus will allow the professionals and experts to take the advantage of the enormous benefits of deep learning models without any hesitation.

\vspace{12pt}
\color{red}


\begin{thebibliography}{00}

\bibitem{optical_co}D. Huang, E. Swanson, C. Lin, J. Schuman, W. Stinson, W. Chang, M. Hee, T. Flotte, K. Gregory, C. Puliafito, and J. Fujimoto: Optical coherence tomography. Science,
254, 1178–1181 (1991) DOI: 10.1126/science.1957169.

\bibitem{intro_pic} Kermany DS, Goldbaum M, Cai W, Valentim CC, Liang H, Baxter SL, McKeown A, Yang G, Wu X, Yan F, Dong J. Identifying medical diagnoses and treatable diseases by image-based deep learning. Cell. 2018 Feb 22;172(5):1122-31.


\bibitem{lee}C. S. Lee, D. M. Baughman, and A. Y. Lee, “Deep learning is effective for classifying normal versus age-related macular degeneration oct images,” Ophthalmology Retina, vol. 1, no. 4, pp. 322–327, 2017.

\bibitem{awais}M. Awais, H. Muller, T. B. Tang, and F. Meriaudeau, “Classification ¨of sd-oct images using a deep learning approach,” in 2017 IEEE International Conference on Signal and Image Processing Applications (ICSIPA). IEEE, 2017, pp. 489–492.

\bibitem{relatedworkresnet}K. He, X. Zhang, S. Ren, and J. Sun, “Deep residual learning for image recognition,” in Proceedings of the IEEE conference on computer vision and pattern recognition, 2016, pp. 770–778.

\bibitem{human_expart}D. S. Kermany, M. Goldbaum, W. Cai, C. C. Valentim, H. Liang, S. L.
Baxter, A. McKeown, G. Yang, X. Wu, F. Yan et al., “Identifying
medical diagnoses and treatable diseases by image-based deep learning,”
Cell, vol. 172, no. 5, pp. 1122–1131, 2018.

\bibitem{kepp2020segmentation}Kepp T, Sudkamp H, von der Burchard C, Schenke H, Koch P, Hüttmann G, Roider J, Heinrich MP, Handels H. Segmentation of retinal low-cost optical coherence tomography images using deep learning. InMedical Imaging 2020: Computer-Aided Diagnosis 2020 Mar 16 (Vol. 11314, p. 113141O). International Society for Optics and Photonics.

\bibitem{wang2019deep}Wang J, Deng G, Li W, Chen Y, Gao F, Liu H, He Y, Shi G. Deep learning for quality assessment of retinal OCT images. Biomedical optics express. 2019 Dec 1;10(12):6057-72.

\bibitem{joshi2020explainable}Joshi A, Mishra G, Sivaswamy J. Explainable Disease Classification via weakly-supervised segmentation. InInterpretable and Annotation-Efficient Learning for Medical Image Computing 2020 Oct 4 (pp. 54-62). Springer, Cham.

\bibitem{singh2020quantitative}Singh A, Balaji JJ, Rasheed MA, Jayakumar V, Raman R, Lakshminarayanan V. Quantitative and Qualitative Evaluation of Explainable Deep Learning Methods for Ophthalmic Diagnosis. arXiv preprint arXiv:2009.12648. 2020 Sep 26.

\bibitem{Kermany2018}Kermany DS, Goldbaum M, Cai W, Valentim CCS, Liang H, Baxter SL, et al. Identifying Medical Diagnoses and Treatable Diseases by Image-Based Deep Learning. Cell. 2018 Feb 22;172(5):1122-1131.e9.

\end{thebibliography}
\end{document}